\documentstyle[12pt,amstex,amssymb,righttag]{article}

\begin{document}
\renewcommand{\thefootnote}{\fnsymbol{footnote}}

\thispagestyle{empty}

\vspace{3cm}

\begin{center}
{\Large \bf New Solutions for Evolution of the Scale Factor with a
Variable Cosmological Term }\\[0.5cm]

\vspace{3mm}
by\\
\vspace{3mm}
{\sl Carlos Pinheiro$^{\ddag}$\footnote{fcpnunes@@cce.ufes.br/maria@@gbl.com.br}
} \\[0.5cm]
and \\[0.5cm]
{\sl F.C. Khanna$^{+}$\footnote{khanna@@phys.ualberta.ca}}

\vspace{3mm}
$^{\ddag}$Universidade Federal do Esp\'{\i}rito Santo, UFES.\\
Centro de Ci\^encias Exatas\\
Av. Fernando Ferrari s/n$^{\underline{0}}$\\
Campus da Goiabeiras 29060-900 Vit\'oria ES -- Brazil.\\

$^{+}$Theoretical Physics Institute, Dept. of Physics\\
University of Alberta,\\
Edmonton, AB T6G2J1, Canada\\
and\\
TRIUMF, 4004, Wesbrook Mall,\\
V6T2A3, Vancouver, BC, Canada.
\end{center}

\vspace{3mm}
\begin{center}
Abstract
\end{center}

The cosmological term is assumed to be a function of time such as
$\Lambda =Ba^{-2}$ where $a(t)$ means the scale factor of standard
cosmology. Analytical solutions for radiation dominated epoch and open
universe are found. For closed universe, $k=+1$, and for flat
universe, $k=0$, we show a numerical solution. In general the scenario
of Big Bang is preserved in our case for the
Friedmann-Robertson-Walker cosmology.

\newpage
\subsection*{Introduction}\setcounter{footnote}{0}

\paragraph*{}
In the last few years several paper \cite{um,cinco} have appeared
discussing  the cosmological constant problem when that term is
considered as part of the content of the universe. In the original
Einstein's proposal the cosmological term was exactly a constant
introduced by hand in Einstein Hilbert lagrangean as a kind of force
opposite to that of gravitation. In the Einstein's equation the
cosmological term appears on the left side and its meaning is only 
geometrical, as must be in the general relativity. However changing
the cosmological constant to the right side in Einstein's equations the
situation is really quite different. Now there is a possibility that
the cosmological term may be time dependent. 

In the new context the
cosmological term could be a function of time with large
possibilities for evolution in time to cosmological term \cite{um}.

The central equation which describes that evolution for scale factor
with cosmological term whether or not it is a function of time is
shown here (7) and elsewhere \cite{um,dois}.

It is possible to find solutions for special cases. In general one can
solve the equation (7) after successive variable transformations on
scale factor variable until the original equation (non linear
differential equation) is written as an ordinary linear differential
equation. However the solutions that are found for almost flat space time,
$k=0$ \cite{um} or under particular hypothesis and numerically are 
approximate solutions.

Here we present some new analytical solutions for the general case,
$k=0$, and $k=\pm 1$. We obtain a direct solution of the equation
without variable transformations for radiation epoch and a numerical
solution for a perfect fluid. 

Differently from \cite{um} we find solutions for flat, open and
closed universes and the big bang scenario is preserved 
\cite{dois}. We consider the usual assumption for a 
Friedman-Robertson-Walker universe like a homogeneous and isotropic universe.

Such a universe is seen as a perfect fluid with a pressure $P$ and
energy density $\rho$.

The Einstein's equations are given by 
\begin{equation}
G_{\mu\nu}=8\pi G\tilde{T}_{\mu\nu}
\end{equation}
where $\tilde{T}^{\mu\nu}$ is the energy-momentum tensor written as 
\begin{equation}
\tilde{T}_{\mu\nu}=T_{\mu\nu}-\frac{\Lambda}{8\pi G}\ g_{\mu\nu}
\end{equation}
It is clear that  $\Lambda$ is a part of the matter content of the universe in
our case. On the right side, $T_{\mu\nu}$ is the usual energy momentum
tensor for perfect fluid and $G$ is the gravitational constant.

The effective energy-momentum tensor describes a perfect fluid and
thus give an effective pressure
\begin{equation}
\tilde{p}=p-\frac{\Lambda}{8\pi G}
\end{equation}
and an effective density as 
\begin{equation}
\tilde{\rho}=\rho+\frac{\Lambda}{8\pi G}
\end{equation}

Then $\tilde{T}_{\mu\nu}$ satisfies energy  momentum conservartion
\cite{tres,quatro} 
\begin{equation}
\nabla^{\nu}\tilde{T}_{\mu\nu}=0
\end{equation}
The equation of state is written as 
\begin{equation}
P=(\gamma -1)\rho
\end{equation}
where $\gamma$ is a constant.

Now, considering the initial hypothesis, eq. (1) and (2), the
Robertson-Walker line element and the fact that the cosmological term
here is not a constant but a function of time, one can show \cite{um}
that the equation which governs the behavior of the scale factor in
the presence of a cosmological term, $\Lambda$, a constant like in the
original Einstein model is given by.
\begin{equation}
\frac{\ddot{a}}{a}=\left(1-\frac{3\gamma}{2}\right)
\left(\frac{\dot{a}^2}{a^2}+\frac{k}{a^2}\right)+
\frac{\gamma}{2}\ \Lambda
\end{equation}
There are numerous possibilites to choose the evolution low for the
cosmological term \cite{um,quatro}. In accordance with \cite{dois} we
choose $\Lambda =\Lambda (a)$, where ``$a$'' means the scale factor for
Friedmann-Robertson-Walker universe. Thus the cosmological term is
written as
\begin{equation}
\Lambda (a)=Ba^{-2}
\end{equation}
where $B$ is a pure number of the order $1$ and ``$a$'' is a function of time.

By combining eq. (7) and (8) we get
\begin{equation}
a\ \frac{d^2a}{dt^2}+\alpha \left(\frac{da}{dt}\right)^2+\lambda =0
\end{equation}
where 
\begin{eqnarray}
\alpha &=&-\left(1-\frac{3\gamma}{2}\right)\quad \mbox{and}\nonumber \\
&&\\
\lambda &=& - \left(1-\frac{3\gamma}{2}\right)k+\frac{\gamma
B}{2}\nonumber 
\end{eqnarray}
There are three free parameters $\gamma$, $k$ and $B$ to be
fixed. The $B$ parameter may be fixed to be $1$ based on dimensional analysis
arguments \cite{dois}.

The parameter $k$, the three spatial curvature assumes the values
$+1$, $0$ and $-1$ for closed, flat and open universe respectively.

\begin{center}
\underline{{\bf Some solutions for equation (9)}}
\end{center}

The dynamic equation for the scale factor $a(t)$ have solutions for
particular values of $\alpha$. 

On taking $\alpha =-1$ the following solutions is found.
\begin{eqnarray}
a(t) &=& \frac{\sqrt{\lambda}\ sinh [c(t-t_0)]}{c}\nonumber \\
&&\\
a(t) &=& \frac{\sqrt{\lambda}\ cosh [c(t-t_0)]}{c}\nonumber 
\end{eqnarray}
where $c$ is a constant.

One fixing $\alpha$ it is easy to verify that $\gamma =0$, and
because of eq. (6) and equation for $\lambda$ we obtain 
\begin{equation}
p=-\rho
\end{equation}
and
\begin{equation}
\lambda =-k\ .
\end{equation}
On taking now $\alpha =-1/2$ the solution is written as 
\begin{equation}
a(t)=-\frac{\lambda}{2c}+c(t-t_0)^2
\end{equation}
and $c$ is a constant.

In this case we find $\gamma =1/3$. The state equation is givne now by
\begin{equation}
p =-\frac{2}{3}\ \rho 
\end{equation}
and
\begin{equation}
\lambda =\frac{-3k+1}{6}
\end{equation}
Next for $\alpha =-2$ the solutions are given by
Jacoby elliptic functions
\begin{eqnarray}
a(t) &=& \sqrt{\frac{\lambda}{2}}\ 
\frac{S_n\left[c(t-t_0),\ \frac{1}{2}\right] 
d_n\left[c(t-t_0),\ \frac{1}{2}\right]}
{c\ c_n\left[c(t-t_0),\ \frac{1}{2}\right]}\ , \nonumber \\
&&\nonumber \\
a(t) &=& \sqrt{-2\lambda}\ 
\frac{d_s\left[c_1(t-t_0),
\frac{1}{2}\right]}{c_1}\\
&&\nonumber \\
\hspace*{-3cm}\mbox{and}\hspace*{3cm} \qquad a(t) &=& \frac{\sqrt{-\lambda}}{c_1\left[c_n(t-t_0),\
\frac{1}{2}\right]}\nonumber 
\end{eqnarray}
where $c,\ c_1$ are constants and $S_n,\ d_n\ c_n$ and $d_s$
are two argument elliptic functions.

In this case we find that $\gamma =-2/3$ and again the equation
of state and $\lambda$ are written as 
\begin{eqnarray}
P &=& -\frac{5}{3}\ \rho \ ,  \\
\lambda &=& -\frac{1}{3} \ (6k+1)\ .
\end{eqnarray}
Consider now the case $\alpha =1$. The solution is found to be
\begin{equation}
a(t)=\pm \sqrt{c-\lambda (t-t_0)^2}
\end{equation}
where $c$ is a constant.

For this case we have $\gamma =4/3$ and the equation for pressure $p$
and $\lambda$ parameter are shown respectively as 
\begin{eqnarray}
p &=& \frac{1}{3}\ \rho \ , \\
\lambda &=& k+\frac{2}{3} \ .
\end{eqnarray}
The following gives the case for $\alpha =-3/2$. The solution is a three
argument Weierstrass function written as.
\begin{equation}
a(t)=\rho \left(c_1(t-t_0)\ , \ \ 0\ \ , \ \ 
-\ \frac{2}{3}\ \frac{\lambda}{c_1^2}\right)
\end{equation}
where $c_1$ is a constant. The same way using eq. (10) and eq. (6) we can
find immediately 
\begin{eqnarray}
p &=& \left(-\frac{4}{3}\right)\ \rho \ , \\
\lambda &=& -\left(2k+\frac{1}{6}\right)\ . 
\end{eqnarray}
Finally, we shall consider the case $\alpha =1/2$. In this case it is
possible to find an analytical solution for the open universe $k=-1$. Using
again eq. (6) and eq. (10) we get 
\begin{eqnarray}
p &=& 0\ , \\
\lambda &=& \frac{1}{2}\ (k+1)\ .
\end{eqnarray}
The solution is shown as
\begin{equation}
a(t)=a_0\left(1+\frac{3}{2}\ \frac{v_0}{a_0}\ (t-t_0)\right)^{2/3}
\end{equation}
with $a_0$ being a constant when we take $a(t)$ at $t=t_0$ and $v_0$
being another constant when $\dot{a}(t)$ is considered at $t=t_0$. The
$dot$ means time derivative.

It's easy to see that for an open universe there is an origen in
time and the big bang scenario is preserved. We find the singularity
$a(t)=0$ as time is expressed as
\begin{equation}
t=t_0-\frac{2}{3}\ \frac{a_0}{V_0}
\end{equation}
For the cases $k=0$ and $k=+1$ we do not find analytical solutions
but we show a numerical solution\footnote{see the Plot for $\alpha =1/2$ 
and $k=0,+1$ at the end of the paper}. It is clear that all
possibilities have an 
origin in time exactly the same way as the standard cosmology where
the cosmological term is a constant.

The Hubble constant may be found as usual i.e.
\begin{equation}
H=\frac{\dot{a}}{a}=H_0\left(\frac{da}{ad\tau}\right)
\end{equation}
where $H_0$ is the present value of Hubble parameter and $\tau =H_0t$
is the measure of time in units of Hubble time \cite{um}.

The analysis of the universe's age follows as usual \cite{dois}.

\subsection*{Conclusions and Comments}

\paragraph*{}
We have found direct solutions of a non linear differential equation
which describes the dynamics for scale factor when the cosmological
term is a function of time. In particular the solution with $\alpha
=1$ describing the radiation dominated epoch and $\alpha =1/2$ describing
a perfect fluid for open universe both preserve the scenario of the
standard cosmology. The possibility that cosmological term is a time
varying function does not change the usual picture of the Friedmann
cosmology when we choose the evolution law eq. (8), except for a
different choice for evolution of $\Lambda$ the final evolution of
the universe can be
different from the usual model. 

The solutions for $\alpha =-1$, $-1/2$, $-2$, $-3/2$ all  have an
appropriete equation of state for vacuum energy for any value of $k$
but these may not be physical.

Finally the case $\alpha =1/2$ for $k=0$, $+1$ is plotted  in figure
I. Clearly we have the big bang scenario for both cases. Only with
our initial hypothesis, the equation of state, eq. (6), and the dynamical
equation for $\Lambda$, the complete behaviour of our model is
similar to the Friedmann cosmology. The analysis of the
age of the universe follows the same steps as in \cite{dois}. The analyse of this problem with 
$\Lambda =Ba^m$ for $m\neq 2$ is presently being considered for the
same initial hypothesis and the equation of state will be obtained.

\subsection*{Acknowledgements:}

\paragraph*{}
I would like to thank the Department of Physics, University of
Alberta for their hospitality. This work was supported by CNPq
(Governamental Brazilian Agencie for Research.

I would like to thank also Dr. Don N. Page for his kindness and attention
with  me at Univertsity of Alberta and Dr. Robert Teshima, Programmer
Analyst from University of Alberta.

%\newpage
%\centerline{\psfig{figure=fig.eps,height=20cm}}
\end{document}